\pdfoutput=1 
\documentclass[iop,author-year]{emulateapj}
\newcommand\icarus{Icarus}%

\usepackage{amssymb}
\usepackage{mathrsfs}
\usepackage{epstopdf}

\newcommand{\fn}[1]{\footnote{\scriptsize{#1}}} 

\newcommand{\Eqn}[1]{Eq{#1}.}  
\newcommand{\Fig}[1]{Fig{#1}.}  
\newcommand{\Voyit}{\textit{Voyager}}  
\newcommand{\NHit}{\textit{New Horizons}}  

\begin{document} 

\title{Compositions and origins of outer planet systems: \\Insights from the Roche critical density} 

\author{Matthew~S.~Tiscareno$^1$, Matthew~M.~Hedman$^1$, Joseph~A.~Burns$^{2,3}$, and Julie Castillo-Rogez$^4$}

\affil{$^1$Center for Radiophysics and Space Research, Cornell University, Ithaca, NY 14853, USA.\\$^2$Department of Astronomy, Cornell University, Ithaca, NY 14853, USA.\\$^3$College of Engineering, Cornell University, Ithaca, NY 14853, USA.\\$^4$Jet Propulsion Laboratory, Pasadena, CA 91109, USA.\\ 
}

\begin{abstract}
We consider the Roche critical density ($\rho_{\mathrm{Roche}}$), the minimum density of an orbiting object that, at a given distance from its planet, is able to hold itself together by self-gravity.  It is directly related to the more familiar ``Roche limit,'' the distance from a planet at which a strengthless orbiting object of given density is pulled apart by tides.  The presence of a substantial ring requires that transient clumps have an internal density less than $\rho_{\mathrm{Roche}}$.  Conversely, in the presence of abundant material for accretion, an orbiting object with density greater than $\rho_{\mathrm{Roche}}$ will grow.

Comparing the $\rho_{\mathrm{Roche}}$ values at which the Saturn and Uranus systems transition rapidly from disruption-dominated (rings) to accretion-dominated (moons), we infer that the material composing Uranus' rings is likely more rocky, as well as less porous, than that composing Saturn's rings.  

From the high values of $\rho_{\mathrm{Roche}}$ at the innermost ring-moons of Jupiter and Neptune, we infer that those moons may be composed of denser material than expected, or more likely that they are interlopers that formed farther from their planets and have since migrated inward, now being held together by internal material strength.

Finally, the ``Portia group'' of eight closely-packed Uranian moons has an overall surface density similar to that of Saturn's A~ring.  Thus, it can be seen as an \textit{accretion-dominated ring system}, of similar character to the standard ring systems except that its material has a characteristic density greater than the local $\rho_{\mathrm{Roche}}$.  

\end{abstract}

\section{The Roche limit and the \\Roche critical density}

The ``Roche limit'' is the distance from a planet within which its tides can pull apart a strengthless compact object.  Simply speaking, a ring\fn{Throughout this work, we use the word ``ring'' to refer to a substantial annulus with sufficient material to support accretion.  Our analysis does not apply to tenuous structures such as Saturn's G~and E~rings.} would be expected to reside inside its planet's Roche limit, while any disk of material beyond that distance would be expected to accrete into one or more moons.  However, the Roche limit does not actually have a single value, but depends particularly on the density, rotation, and internal material strength of the moon that may or may not get pulled apart \citep{Weidenschilling84,CE95}.  A simple value for the Roche limit of a spherical moon, assuming no internal strength, can be calculated from a balance between the tidal force (i.e., the difference between the planet's gravitational pull on one side of the moon and its pull on the other side) that would tend to pull a moon apart, and the moon's own gravity that would tend to hold it together.  This gives \citep[e.g., \Eqn{}~4.131 in][]{MD99}
\begin{equation}
\label{RocheLimitEqn}
a_{\mathrm{Roche}} = R_{\mathrm{p}} \left( \frac{4 \pi \rho_\mathrm{p}}{\gamma \rho} \right)^{1/3} , 
\end{equation}
where $R$ is radius and $\rho$ is internal density, and the subscript ``p'' denotes the central planet.   

The dimensionless geometrical parameter $\gamma = 4 \pi / 3 \approx 4.2$ for a sphere, but is smaller for an object that takes a non-spherical shape with its long axis pointing toward Saturn, as one would expect for an actively accreting body and as observed for several of Saturn's ring-moons \citep{PorcoSci07,Charnoz07}.  The region of gravitational dominance of a point mass moon, known as the ``Roche lobe'' or the ``Hill sphere,'' is lemon-shaped, with cusps at the two Lagrange points $L_1$ and $L_2$ \citep[see, e.g., \Fig{}~3.28 of][]{MD99}.  Simply distributing the moon's material into the shape of its Roche lobe, with uniform density, yields $\gamma \approx 1.6$ \citep{PorcoSci07}.  Going further for the case of an incompressible fluid, fully accounting for the feedback between the moon's distorted shape and its (now non-point-mass) gravity field smooths out the cusps and further elongates the moon \citep{Chandra69,MD99}, leading to an even smaller value, $\gamma \approx 0.85$.  However, some central mass concentration and the inability of a rubble pile with internal friction to exactly take its equilibrium shape will likely prevent $\gamma$ from becoming quite this low.  Hereafter, we will use the \citet{PorcoSci07} value of $\gamma \approx 1.6$. 

\begin{figure}[!t]
\begin{center}
\includegraphics[width=8cm]{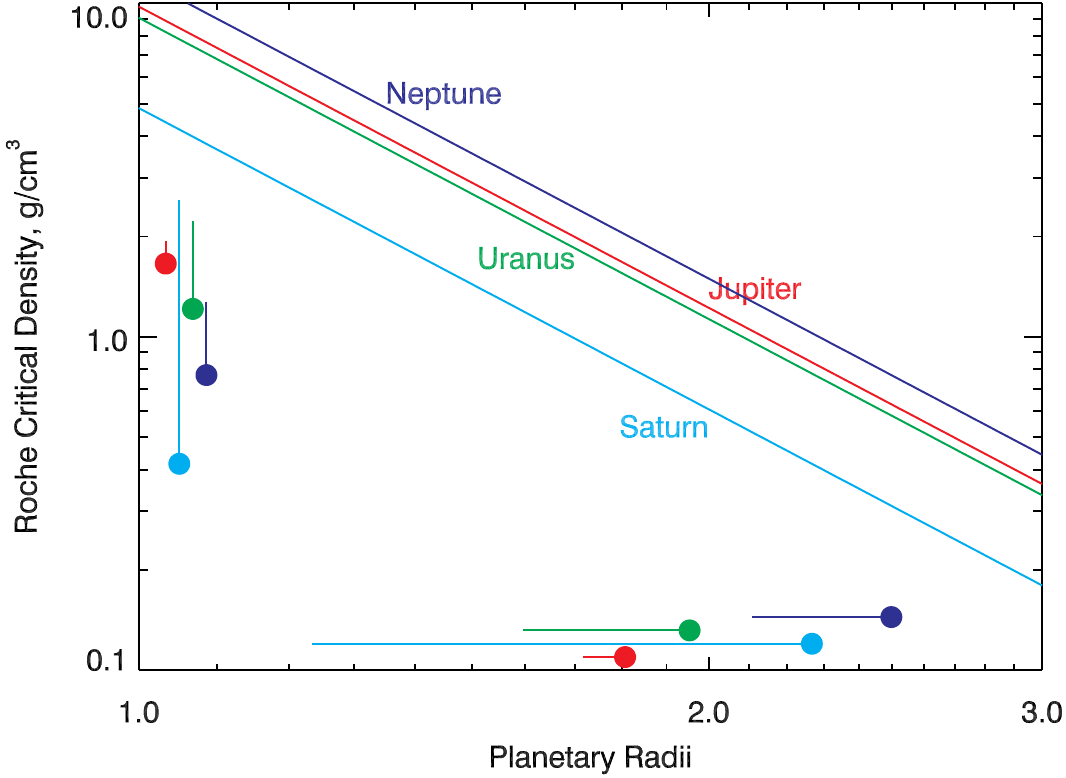}
\caption{The Roche critical density $\rho_{\mathrm{Roche}}$ (\Eqn{}~\ref{RocheDensEqn1} or~\ref{RocheDensEqn2}, with $\gamma = 1.6$) plotted against planetary radii for Jupiter (red), Saturn (cyan), Uranus (green), and Neptune (blue).  An object must have a density higher than $\rho_{\mathrm{Roche}}$ to be held together by its own gravity; conversely, in the presence of abundant disk material, an embedded object will actively accrete as long as its density remains higher than $\rho_{\mathrm{Roche}}$.  The colored bars along the bottom show the extent of each planet's main ring system.  The bars along the left-hand side are constructed from the bars along the bottom simply by the latter reflecting off the appropriately colored diagonal line.  For each bar, a solid circle indicates the outermost extent, and the corresponding minimum $\rho_\mathrm{Roche}$, of the main rings.  Figure from \citet{Ringschapter13}.
\label{rochedens}}
\end{center}
\end{figure}

\begin{figure*}[!t]
\begin{center}
\includegraphics[width=14cm]{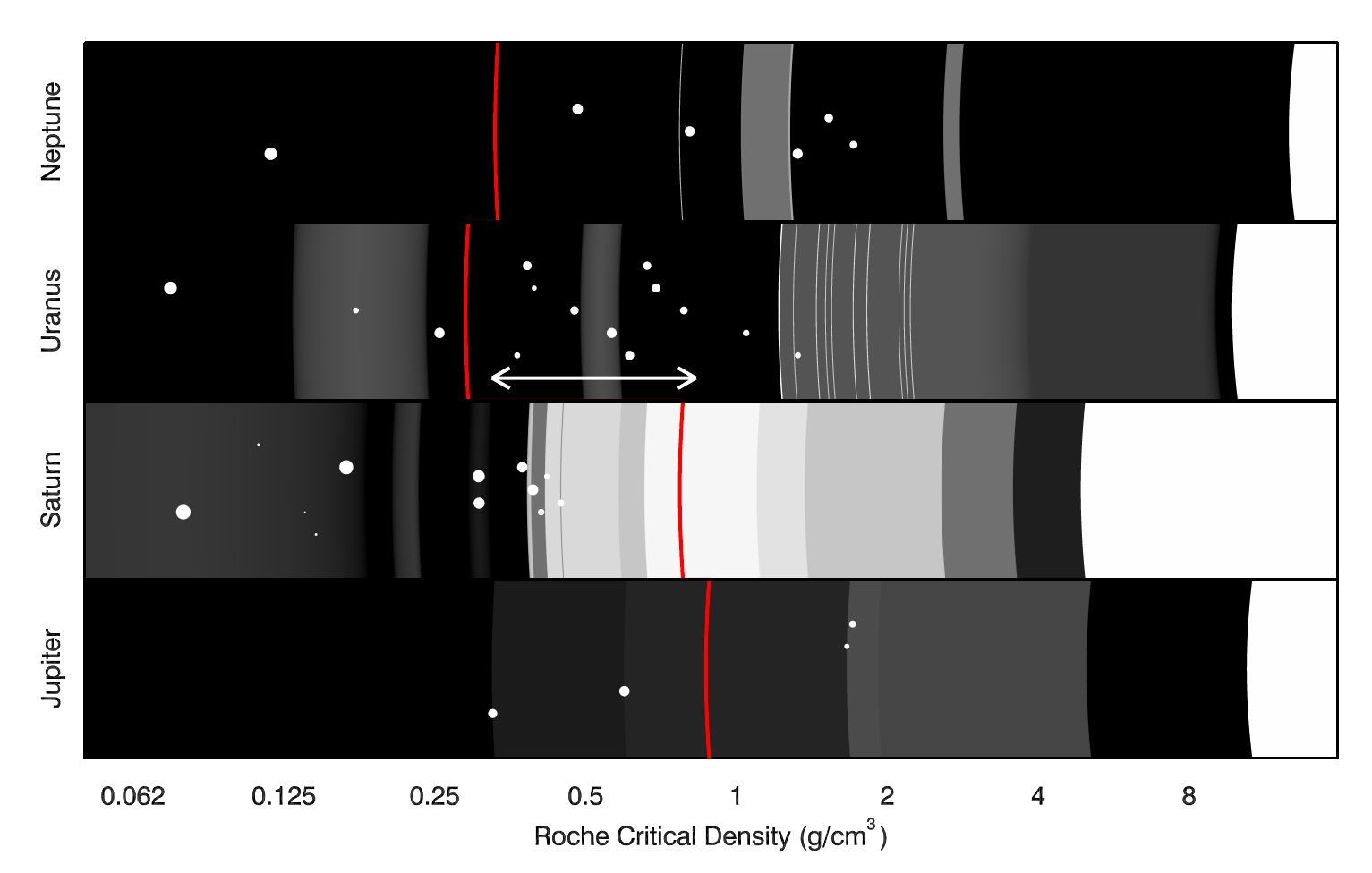}
\caption{Schematic of the four outer planet ring systems, with a common scale based on $\rho_{\mathrm{Roche}}$.  Shades of gray indicate ring surface densities, with black indicating regions that are empty (so far as is known) and white indicating the highest surface densities (or the planet itself).  The red line in each panel indicates the location where the orbit rate is synchronous with the planet's rotation.  In the Uranus panel, the arrow indicates the extent of the ``Portia group'' of moons, as discussed in the text.  Figure from \citet{HedmanChapter13}.  
\label{HedmanRocheFig}}
\end{center}
\end{figure*}

Because the moon's internal density $\rho$ appears in \Eqn{}~\ref{RocheLimitEqn}, there is no single value of the Roche limit for a planetary system.  This is intuitive, as a denser object can venture closer to the planet without danger of fragmentation than can an object that is less dense \citep[see \Fig{}~1 of][]{Weidenschilling84}.  In fact, it is often more useful in the context of rings to consider the limit from planetary tides not as a critical distance but rather as a critical density.  

We can rearrange \Eqn{}~\ref{RocheLimitEqn} such that, at any given distance $a$ from the planet, the Roche critical density $\rho_{\mathrm{Roche}}$ at which the moon's size entirely fills its Roche lobe is  
\begin{equation}
\label{RocheDensEqn1}
\rho_{\mathrm{Roche}} = \frac{4 \pi \rho_\mathrm{p}}{\gamma (a/R_{\mathrm{p}})^3} . 
\end{equation}
Then, substituting the planet's mass $M_\mathrm{p} = (4 \pi / 3) R_\mathrm{p}^3 \rho_\mathrm{p}$, we have
\begin{equation}
\label{RocheDensEqn2}
\rho_{\mathrm{Roche}} = \frac{3 M_{\mathrm{p}}}{\gamma a^3} . 
\end{equation}
While \Eqn{}~\ref{RocheDensEqn2} gives the simplest method for calculating $\rho_\mathrm{Roche}$, \Eqn{}~\ref{RocheDensEqn1} expresses its dependence on the distance in terms of planetary radii ($a/R_\mathrm{p}$); when shown on a log-log plot, as in \Fig{}~\ref{rochedens}, it appears as a straight line of slope $-3$, normalized by each planet's internal density $\rho_\mathrm{p}$. 

Within a ring, where material for accretion is plentiful and collisions are relatively gentle, any pre-existing solid chunk with internal density $\rho > \rho_{\mathrm{Roche}}$ should accrete a mantle of porous ring material until its density decreases to match $\rho_{\mathrm{Roche}}$ (which is to say, the resulting entity fills its Roche lobe).  Indeed, the moons near and within Saturn's rings are observed to have $\rho \sim \rho_\mathrm{Roche}$ and to have shapes reminiscent of their Roche lobes \citep{PorcoSci07,Charnoz07}.  On the other hand, any loose agglomeration of material with internal density  $\rho < \rho_{\mathrm{Roche}}$ must shed material.  Such an object may be the aforementioned solid core with an icy mantle, if it has grown beyond its Roche lobe, in which case it will simply shrink back to the size of its Roche lobe.  However, a simple clump of material with no core cannot increase its internal density by shedding material, and thus must disjoin completely if its $\rho$ is less than $\rho_{\mathrm{Roche}}$ by a sufficient margin.\fn{Even a rubble pile has some internal material strength induced by granular friction, which can serve to hold it together in the face of tidal forces that would otherwise tear it apart.  Therefore, the actual critical density for breakup is somewhat less than $\rho_\mathrm{Roche}$ \citep[e.g.,][]{Sharma09}.  In this work, however, we assume that this effect is small and do not treat it quantitatively.}  Therefore, the fundamental criterion determining whether a particular location in a planetary system is characterized by disruption (rings) or accretion (discrete moons) is whether the density naturally achieved by transient clumps is greater or less than $\rho_{\mathrm{Roche}}$.

\section{Applications}

\subsection{A clue to composition}
As explained in the previous section, the persistent existence of a ring at a given radial location $a$ implies that the densities of transient clumps $\rho_\mathrm{clump}$ do not exceed $\rho_{\mathrm{Roche}}$ --- that is, we expect that the composition of the ring material is such that $\rho_\mathrm{clump} \lesssim \rho_{\mathrm{Roche}}$.  Because of the porosity inherent in any transient clump,\fn{Recall that the volume fraction of close-packed equal spheres is only $\sim 70$\%.} we can expect $\rho_\mathrm{clump}$ to be only a fraction of $\rho_\mathrm{solid}$, the density of a solid chunk of the material that composes the ring.  

As seen in \Fig{}~\ref{rochedens}, Saturn's main rings extend outward to significantly lower values of $\rho_{\mathrm{Roche}}$, approaching 0.4~g~cm$^{-3}$, than are seen in any of the other three known ring systems, probably reflecting their much lower rock fraction (and higher fraction of water ice) as already known from spectroscopy and photometry.  In fact, Saturn's rings are composed almost entirely of water ice \citep{CuzziChapter09}, so we can conclude that $\rho_\mathrm{clump}$ is $\sim 40\%$ of $\rho_\mathrm{solid}$, at least in this case.  Saturn's ``ring moons'' from Pan through Pandora, which range from 4~to 40~km in mean radius and orbit in the vicinity of the rings, have similar densities and porosities \citep{Thomas10}. 

The compositions of the Uranian and Neptunian rings are almost entirely unknown, as \Voyit{} did not carry an infrared spectrometer with enough spatial resolution to detect them.  However, it is clear from their low albedo that at least the surfaces of the ring particles cannot be primarily water ice.  Color imaging indicates that the Uranian rings are dark at all visible wavelengths, which would be consistent with the spectrum of carbon or organics, among other possibilities. 

Like at Saturn, the Uranus system has a clear boundary between disruption-dominated and accretion-dominated regions (\Fig{}~\ref{HedmanRocheFig}), notwithstanding some minor mixing at the boundary --- at Uranus, Cordelia is inward of the $\epsilon$~ring; at Saturn, Pan and Daphnis orbit within the outermost parts of the A~ring.  The Uranian main rings extend only to\fn{The limiting value of $\rho_{\mathrm{Roche}}$ for the Uranus system may be as high as 1.4~g~cm$^{-3}$ (the value for the $\delta$~ring) 
if the highly perturbed state of the $\epsilon$~ring increases its resistance to accretion, a question that has not been studied in detail.} $\rho_{\mathrm{Roche}} \approx 1.2$~g~cm$^{-3}$.  If we assume a porosity for Uranian ring material with a similar value ($\sim 40\%$) as at Saturn, then a solid chunk of the material that composes the ring could be as high as $\rho_\mathrm{solid} \approx 3$~g~cm$^{-3}$. 
However, it is probably not that high, as that value would imply a composition almost entirely of rock, while lower-density materials (such as ices, clathrates, and perhaps organics) are likely to be present in significant amounts.  Nonetheless, we can safely conclude from this observation that the Uranian rings are likely composed of material with a much higher rock fraction than at Saturn, as well as that the porosity of transient clumps is somewhat less than at Saturn. 

Our inference from the Roche critical density at the ring/moon transition, that the Uranus system is rockier overall than the Saturn system, is consistent with the fact that the average density of Saturn's mid-size moons \citep{MatsonChapter09} is 1.2~g~cm$^{-3}$, while that of Uranus' major moons \citep{Jake92} is 1.6~g~cm$^{-3}$ 
(although the lower density of Uranus' innermost major moon, Miranda at $1.2 \pm 0.15$ g/cm$^3$, may hint at heterogeneous accretion within the protosatellite disk).  Furthermore, the 40\% increase in moon density from Saturn to Uranus, rather than a threefold increase as for the transitional $\rho_{\mathrm{Roche}}$, supports the inference that porosity is greater in Saturn's ice-rich ring material. 

The outer edge of Neptune's main ring system corresponds to $\rho_{\mathrm{Roche}} \approx 0.8$~g~cm$^{-3}$.  This is intermediate between the values for the Saturn and Uranus systems, possibly indicating an intermediate rock/ice ratio.  However, Neptune's rings are much less substantial than Saturn's or Uranus' (its highest optical depths are comparable to those of the C~ring and the Cassini Division, but with considerably more dust), and even the Adams ring may be nothing more than an assemblage of dust-shedding moonlets.  Furthermore, any indigenous major moons of Neptune were likely lost during Triton's capture \citep{GoldreichNeptune89,AH06}, so we have no independent way to estimate the composition of Neptune's small moons or rings.  We therefore cannot use the location of Neptune's rings to provide a reliable constraint on the ring material's composition or porosity. 

The extent of Jupiter's Main ring, in contrast to the other three ring systems, is clearly limited by the availability of material (which, other than dust, is restricted to the narrow region between the source moons Metis and Adrastea, and which is not abundant) rather than by a disruption/accretion balance.  However, Jupiter's high transitional $\rho_{\mathrm{Roche}} \approx 1.7$~g~cm$^{-3}$ places the only known limit on the densities (and thus masses) of the source moons, which must be denser than $\rho_{\mathrm{Roche}}$ in order to hold themselves together by gravity.  However, it may not be valid to assume that Metis and Adrastea are held together by gravity, as accreting masses must be, given the large gap in particle size between the $\sim 10$-km moons and other Main ring particles, which observationally cannot be larger than 1~km \citep{Show07}.  This large gap in particle size might be explained if Metis and Adrastea are solid bodies originating further from Jupiter, now held together by material strength, while no bodies of similar size are now able to form through \textit{in~situ} accretion. 

\subsection{A clue to past dynamics}
Unlike Saturn and Uranus, Jupiter and Neptune do not have clear boundaries between accretion-dominated and disruption-dominated regions.  Both have moons interspersed with their rings, down to values of $\rho_{\mathrm{Roche}}$ as high as 2~g~cm$^{-3}$ (\Fig{}~\ref{HedmanRocheFig}).  In the case of Jupiter, the value of $\rho_{\mathrm{Roche}}$ in the vicinity of the innermost moons is twice as high as the measured bulk density of Amalthea \citep{Anderson05}.  In the case of Neptune, it is three times as high as the constraint placed on the densities of the inner moons by \citet{ZH07,ZH08} in order to account for their low inclinations.  The existence in both systems of moons at such high values of $\rho_{\mathrm{Roche}}$ may indicate that they are interlopers that formed farther from their planets and have since migrated inward and are held together by internal material strength.  Alternatively, it may indicate that they are composed of denser (likely silicate) material.  
As the internal strength required to hold together a moon is given by \citep{HM08}
\begin{equation}
\label{StrengthHM}
\bar{\sigma} = \frac{4 \pi R^2 G \rho}{15} \left( \rho_\mathrm{Roche} - \rho \right) , 
\end{equation}
\noindent our analysis indicates that all of the known moons in the outer solar system can be held together with internal strengths of less than 50~kPa (Table~\ref{StrengthTable}), which is several orders of magnitude smaller than typical strengths for ice-rock mixtures \citep{Durham92}. 

\begin{table}
\caption{Internal strength required to hold together the innermost known moon in each of the giant planet systems, if $\rho = 1$~g~cm$^{-3}$. \label{StrengthTable}}
\begin{scriptsize}
\hspace{0.3cm}
\begin{tabular} { l c c c c }
\hline
\hline
Name (System) & $R$ (km) & $\rho_\mathrm{Roche}$ (g~cm$^{-3}$) & $\bar{\sigma}$ (kPa) \\
\hline
Metis (Jupiter) & 22 & 1.71 & 19.2 \\
Pan (Saturn) & 14 & 0.45 & --- \\
Cordelia (Uranus) & 20 & 1.32 & 7.2 \\
Naiad (Neptune) & 33 & 1.71 & 43.4 \\
\hline
\end{tabular}
\end{scriptsize}
\vspace{0.6cm}
\end{table}

In the case of Neptune, the icy composition of most objects that far from the Sun suggests that its moons have significant internal strength, though it is possible for Neptune's moons to be more silicate-rich than its rings (as is, in fact, the case for Saturn's mid-size moons, though not for Saturn's ring-moons).  In the case of Jupiter, the importance of internal strength is corroborated by \NHit{}'s non-detection of objects with sizes intermediate between Metis and Adrastea and the continuum ring particles \citep{Show07}.  

The hypothesis that the inner moons of Jupiter and Neptune may have formed farther from their planets may even include possible formation elsewhere in the solar system.  \citet{CharnozLHB09} found that Jupiter and Neptune are likely to have received the most material from circum-solar orbits during the Late Heavy Bombardment; therefore, those two planets are the most likely to have captured an object from heliocentric orbit.  Capture could be facilitated near the Roche limit (given the captured object's density) as internal dissipation is increased for an object near breakup. 

However, a gentler scenario is available, in that the inner moons of both Neptune and Jupiter (as well as Uranus) are well inward of synchronous orbit (\Fig{}~\ref{HedmanRocheFig}).  So they may have formed in the latter vicinity, where $\rho_{\mathrm{Roche}}$ is relatively low, and subsequently migrated inward under tidal evolution.  Dynamical constraints on how much Io can have migrated over its history \citep{HamiltonDDA11} should not limit the movement of Jupiter's inner moons to their current locations through gradual processes (D.~P.~Hamilton, personal communication,~2012). 

\subsection{An accretion-dominated ring}
Uranus' inner complement of moons (which we will call the ``Portia group'' after its leading member) is the most densely-packed known satellite system, with eight moons occupying an annulus from 2.31~to 2.99~$R_\mathrm{U}$ \citep{DL97,SL06}.  The orbits of the moons in this group are dynamically unstable on timescales of $\sim 10$~Myr \citep{SL06,DawsonDDA10,FS12}, and the region has likely looked quite different over solar system history as moons are disrupted and re-accreted on a regular basis.  The dusty $\nu$~ring, which lies between two of the moons of the Portia group, may indeed be the detritus of a recent significant collision, perhaps the disruption of a moon.  \citet{FS12} have suggested that, some 10~Myr in the future, it is likely that the $\nu$~ring will have re-accreted into a new moon, while one or more current moons (Cupid seems a likely candidate) will have been destroyed and will temporarily look as the $\nu$~ring does now.  However, detailed study of this scenario remains to be carried out. 

The mean surface density of the Portia group region, calculated by spreading the moons' mass evenly over the annulus containing them, is $\sim 45$~g~cm$^{-2}$, 
comparable to the surface density of Saturn's A~ring \citep{soirings}.  With all this in mind, Uranus' Portia group can be seen as being of similar character to the known dense ring systems, with the difference being that it is dominated by accretion rather than by disruption, due to the low value of $\rho_{\mathrm{Roche}}$ at its location. 

\section{Conclusions}
We consider the Roche critical density ($\rho_{\mathrm{Roche}}$), the minimum density of an orbiting object that, at a given distance from its planet, is able to hold itself together by self-gravity.  It is directly related to the more familiar ``Roche limit,'' the distance from a planet at which an orbiting object of given density is pulled apart by tides.  At a given distance from the planet, an orbiting object whose density is less than $\rho_{\mathrm{Roche}}$ will be pulled apart (unless it has internal material strength to hold itself together), and its material will form a ring.  Conversely, in the presence of abundant material for accretion, an orbiting object whose density is greater than $\rho_{\mathrm{Roche}}$ will accrete material; if the accreted material is more fluffy than the dense core, the object's bulk density may decrease until equal to $\rho_{\mathrm{Roche}}$, at which point it will have ``filled its Roche zone'' and will stop accreting.

Both Saturn and Uranus have relatively clear boundaries between accretion-dominated regions (populated with moons) and disruption-dominated regions (populated with rings).  Given the presence, in both cases, of abundant material for accretion, the value of $\rho_{\mathrm{Roche}}$ at the boundary should be slightly higher than the density of transient clumps.  Noting that the boundary value at Uranus is three times higher than at Saturn, we infer that the material composing Uranus' rings is likely more rocky, as well as less porous, than that composing Saturn's rings.  This inference is consistent with several observational clues. 

Neptune and Jupiter have moons and rings interspersed, with moons at values of $\rho_{\mathrm{Roche}}$ as high as 2~g~cm$^{-3}$.  This may indicate that the moons are interlopers that formed farther from their planets and have since migrated inward and are held together by internal material strength, or it may indicate that they are composed of denser (likely silicate) material.

The ``Portia group'' of eight Uranian moons packed between 2.31~to 2.99~$R_\mathrm{U}$ has an overall surface density similar to that of Saturn's A~ring.  Thus, it can be seen as an \textit{accretion-dominated ring system}, of similar character to the standard ring systems except that its material has a characteristic density greater than the local $\rho_{\mathrm{Roche}}$.  This inference is consistent with the high dynamical instability of these moons, as well as with the existence of the $\nu$~ring without any clear extant antecedent. 

\acknowledgements 
We thank John Weiss, Doug Hamilton, Matija \'Cuk, and Rick Greenberg for helpful conversations.  M.S.T. acknowledges funding from the NASA Outer Planets Research program (NNX10AP94G).

\end{document}